\documentstyle[12pt,epsfig,amsfonts]{article} 
\newlength{\dinwidth} 
\newlength{\dinmargin} 
\newcommand{\ba}{\begin{array}} 
\newcommand{\ea}{\end{array}} 
\newcommand{\be}{\begin{equation}} 
\newcommand{\ee}{\end{equation}} 
\newcommand{\bea}{\begin{eqnarray}} 
\newcommand{\eea}{\end{eqnarray}} 
\newcommand{\gsim}{\mathrel{\mathop{\kern 0pt \rlap 
  {\raise.2ex\hbox{$>$}}} \lower.9ex\hbox{\kern-.190em $\sim$}}}

\def\bz{{\bar z}}

\def\log{{\rm{log}}}

\def\d{{\rm d}}

\def\ben{\begin{equation}} 
\def\een{\end{equation}} 
\def\bea{\begin{eqnarray}} 
\def\eea{\end{eqnarray}}

\def\pd{\partial} 
\def\a{\alpha} 
\def\b{\beta} 
\def\g{\gamma} 
\def\d{\delta} 
\def\m{\mu} 
\def\n{\nu} 
\def\t{\tau} 
\def\l{\lambda}

\def\s{\sigma} 
\def\e{\epsilon}

\def\ene{{\cal N}}

\def\defcon{{\cal\beta}^{\mathnormal b}}

\def\entero{\mathbb{Z}}
\def\real{\mathbb{R}}

\begin{document} 
\thispagestyle{empty} 
\addtocounter{page}{-1} 
\vskip-0.35cm 
\begin{flushright}
IFT 04/35\\
{\tt hep-th/0407045}
\end{flushright}
\vspace*{0.2cm}
\centerline{\Large \bf Fluxes, Tadpoles and Holography}
\vskip0.3cm
\centerline{\Large \bf for ${\cal N}=1$ Super Yang Mills}

\vspace*{1.0cm}
\centerline{\bf C\'esar G\'omez, Sergio Monta\~nez and Pedro Resco}
\vspace*{0.7cm}

\centerline{\it Instituto de F\'\i sica Te\'orica CSIC/UAM, C-XVI Universidad Aut\'onoma,}
\vspace*{0.2cm}
\centerline{\it E-28049 Madrid \rm SPAIN}
\vspace*{1cm}
\centerline{\tt cesar.gomez@uam.es }
\vspace*{0.2cm}
\centerline{\tt sergio.montannez@uam.es }
\vspace*{0.2cm}
\centerline{\tt juanpedro.resco@uam.es }

\vspace*{0.8cm}
\centerline{\bf abstract}
\vspace*{0.3cm}

 We study non perturbative superpotentials for $\ene=1$ Super Yang Mills from 
the point of view of large $N$ dualities. Starting with open topological 
strings we work out the relation between the closed string sector 
tadpole and NSNS fluxes in the 
closed string dual on the resolved conifold. For the mirror closed string 
dual version on the deformed conifold we derive, by computing the $G_3$ flux induced superpotential, the $N$ supersymmetric vacua and study the transformations of $G_{3}$ through domain walls. The Wilsonian beta function is discussed in this context. Finally, as an extension of Fischler Susskind 
mechanism we find a relation between the tadpole and the 
geometric warping factors induced by the gravitational backreaction of NSNS 
 fluxes.

\baselineskip=18pt

\newpage

\section{Introduction}

The recent progress in the understanding of large $N$ dualities  \cite{'tHooft:1973jz,Maldacena:1997re,Gopakumar:1998ki} provides important hints for solving the mass gap dynamical generation problem in $\ene=1$ supersymmetric gauge theories, more precisely the gaugino condensate problem. There are two complementary main avenues that have been worked out in the last few years. One based
on holographic duals \cite{Klebanov:2000hb,Maldacena:2000yy,Polchinski:2000uf} where the dynamically generated scale
of $\ene=1$ Super Yang Mills is directly related to the size of some internal target space-time
cycle. The other approach \cite{Vafa:2000wi,Cachazo:2001jy} is based on Gopakumar-Vafa \cite{Gopakumar:1998ki} large $N$ duality. The main point
in this last approach is to use the large $N$ duality between Chern-Simons on $S^{3}$ and
topological closed strings on the resolved conifold as a tool to define the non perturbative contributions to the $\ene=1$ Super Yang-Mills superpotential \cite{Veneziano:1982ah}.

From the brane-flux interpretation of open-closed dualities it is natural
to expect that the closed string dual is defined by some target geometry in
addition to some RR and NSNS fluxes. In particular in our case of interest we will deal with the resolved conifold with extra RR and NSNS fluxes \cite{Vafa:2000wi}. Our first interest in this paper
is to understand the open string ancestor of the NSNS flux. We find that this flux
is directly related to the {\em closed string sector tadpole} associated with
the annulus amplitude of the open topological string.

This kind of tadpoles has been considered in critical string theories from the point of view of Fischler-Susskind \cite{Fischler:ci} mechanism as a way to modify the target space-time. However we can also consider this mechanism from a {\em holographic  (or non critical) point of view}. As we will show, we can directly read the geometric warping factor induced by the gravitational back reaction of the NSNS flux from the tadpole, which is a different manifestation on the well known relation between NSNS fluxes and geometrical warping factors \cite{Giddings:2001yu}.

The paper is organized as follows. In section 2 we first review the connection between the non perturbative free energy of Chern-Simons field theory on $S^3$ and the non perturbative contributions for the $\ene=1$ superpotential. We find that the complete derivation of the Veneziano-Yankielowicz superpotential for $\ene=1$ requires to take into account both the non perturbative contribution to the Chern-Simons genus zero free energy and also the annulus tadpole of the open topological string. Once we have both ingredients we can identify the dynamically generated scale of $\ene=1$ Super Yang-Mills with the IR cutoff appearing in the computation of the annulus tadpole. In section 3 we discuss the superpotential $W$ from the closed string point of view on the mirror $B$ model as a function of the complex structure parameter $t$ of the $CY_3$, deriving the expected $N$ supersymmetric vacua of $\ene=1$ Super Yang-Mills and the corresponding domain wals. Finally we discuss the derivation of beta functions.

\section{The Gravity Dual of $\ene=1$ Super Yang-Mills: The Topological Approach}

The topological approach \cite{Vafa:2000wi} to the construction of a closed string gravity dual of $\ene=1$ Super Yang-Mills is based on the following facts:
\begin{itemize}
\item[i)] Type A topological open string amplitudes on the Calabi-Yau manifold $T^*S^3$ induce a F-term superpotential $W(S)$, with $S= trW^2$ the glueball superfield, of four dimensional $\ene=1$ SYM gauge theory \cite{Bershadsky:1993cx,Antoniadis:1993ze}. More concretely
\ben\label{one} 
\int d^{2}\theta W(S) = \int d^{2}\theta \sum_{h} F^{(s)}_{0,h} N h S^{h-1}\equiv N \int d^{2}\theta \frac{\pd F^{(s)}_0}{\pd S}
\een
where $F^{(s)}_{0,h}$ are open string amplitudes at genus zero and with $h$ loops with Dirichlet boundary conditions on the lagrangian submanifold $S^3$, and where we define $F_0^{(s)}(S)=\sum_{h} F^{(s)}_{0,h} S^{h}$.
\item[ii)] Type A topological open string theory on $T^*S^3$ is equivalent to three dimensional Chern-Simons gauge theory on $S^3$ \cite{Witten:1992fb}. This means that, for the Chern-Simons free energy
\be
\label{two}
F^{(CS)}=\sum_g F_g^{(CS)}(\lambda) \left( \frac{\lambda}{N}  \right)^{2g-2}
\ee
with
\be
\label{three}
 F_g^{(CS)}(\lambda)=\sum_h C_{g,h}\lambda^h
\ee
and with $\lambda$ the 't Hooft coupling constant $\lambda=\frac{2\pi N}{k+N}$, being $k$ the level, we have
\be
\label{four}
C_{g,h}=F^{(s)}_{g,h}
\ee
Using (\ref{one}), (\ref{three}) and (\ref{four}) we can write
\be
\label{five}
W(S)=N \frac{\pd F_0^{(CS)}(\lambda)}{\pd \lambda} { \Bigg \arrowvert}_{\lambda=S}
\ee
A non perturbative contribution to $W(S)$ in the regime $\lambda=S$ small can be directly obtained using (\ref{five}) and the exact solution for Chern-Simons theory \cite{Witten:1988hf,Ooguri:2002gx,Periwal:1993yu}
\be
\label{six}
W^{np}(S)=N \frac{\pd F_{0(np)}^{(CS)}(\lambda)}{\pd \lambda} { \Bigg \arrowvert}_{\lambda=S}
\ee
where
\be
\label{seven}
 F_{0(np)}^{(CS)}(\lambda)=\frac{\lambda^2}{2}\left(\log \lambda -\frac{3}{2}  \right)
\ee
This formally produce a  Veneziano-Yankielowicz (VY) \cite{Veneziano:1982ah} type of superpotential for $S$:
\be
\label{eight}
W^{np}=NS \left[ \log S-1  \right]
\ee
\item[iii)] A the necessary step in order to construct a closed string gravity dual is to use the Gopakumar-Vafa large $N$ duality between Chern-Simons on $S^3$ and type A topological closed strings on the resolved conifold \cite{Gopakumar:1998ki}, which proposes the equivalence between both theories provided that the imaginary part of the complexified K\"ahler modulus of the conifold takes the value
\be
t=\lambda
\ee
\item[iv)] It was shown in \cite{Bershadsky:1993cx,Antoniadis:1993ze} that the closed type A topological string amplitudes $F_{g}(t)$ on the resolved conifold $X$ compute a four dimensional effective supergravity
lagrangian for the corresponding compactification of type IIA string theory. From this lagrangian, one can see that, if there are also $N$ units of RR flux, $F_{0}(t)$ computes 
an effective $\ene=1$ superpotential for the moduli superfield $t$. This closed string dual version of the non perturbative superpotential (\ref{eight}) is the Gukov-Vafa-Witten (GVW) superpotential \cite{Taylor:1999ii,Gukov:1999gr,Gukov:1999ya,Mayr:2000hh}:
\be
\label{nine}
W(t)=N \frac{\pd F_0}{\pd t}
\ee
where $F_0(t)$ is the closed topological string amplitude at genus zero on the resolved conifold before adding fluxes. Notice that (\ref{nine}) is correct only if topological amplitudes are unaffected by the gravitational back reaction of RR-fluxes (see appendix A for a short discussion on this result).
\end{itemize}

From i), ii), iii) and iv) we can say that concerning F-terms the gravity dual of $\ene=1$ SYM is just type IIA closed superstring theory on the resolved conifold with fluxes turned on and with
\be
\label{Vafaiden}
t=<S>
\ee
There are however some details in this derivation on which we want to focus our attention on the rest of this section.

\subsection{Scales, Renormalization Group Invariance and NSNS Fluxes}

First of all, in order to get $W^{np}(S)$ in (\ref{eight}) we have used the formal identification $S= \lambda$ for $\lambda$ the 't Hooft coupling of Chern-Simons theory. In order to make this identification consistent with dimensions we should introduce some arbitrary energy scale $\mu$ and to define
\be
\label{ten}
W^{np}(S)=\mu^3 W^{np}(\lambda)
\ee
for $\lambda=\frac{S}{\mu^3}$. The way we introduce the scale $\mu$ in this context is exactly the same used in standard quantum field theory in dimensional regularization \cite{'tHooft:1972fi}. From (\ref{ten}) we get
\be
\label{eleven}
W^{np}(S)=NS \left[ \log \left( \frac{S}{\mu^3}  \right)-1  \right]
\ee
The first problem with (\ref{eleven}) is obviously that it is not RG invariant with respect to changes of $\mu$. The way this problem is solved for VY superpotential is in fact well known. You simply define
\be
\label{twelve}
W^{VY}(S)=NS \left[ \log \left( \frac{S}{\mu^3}  \right)-1  \right]+\tau(\mu)S
\ee
with $\tau(\mu)=\frac{8\pi^2}{g_{YM}^2(\mu)}+i\theta_{YM}$ is the $\ene=1$ SYM holomorphic (Wilsonian) coupling at scale $\mu$. In fact, using the exact holomorphic beta function \cite{Arkani-Hamed:1997mj}
\be
\label{exactbeta}
\beta=\mu \frac{\pd g_{YM}}{\pd \mu}=-\frac{3Ng_{YM}^3}{16 \pi^2}
\ee 
one can easily prove that (\ref{twelve}) is RG invariant.

From the perspective of the closed string gravity dual it is natural to look for a GVW superpotential reproducing (\ref{twelve}). The answer suggested in \cite{Vafa:2000wi} is that the extra term in (\ref{twelve}) comes from an extra NSNS 4-form flux. Of course, this extra NSNS flux is going to produce a gravitational back reaction that very likely is going to change the background geometry. However, since we are interested in reproducing (\ref{twelve}), we maintain unaffected the topological closed string amplitude $F_0(t)$ from which we have derived the first term in (\ref{twelve}). Again, a discussion on this result is given in the appendix A. Here we want to stress that RG invariance with respect to changes in $\mu$, from which we can read beta functions, only depends, at this level of discussion, on the form of the superpotential that is not taking into account the gravitational back reaction of NSNS fluxes. Of course, an holographic interpretation of the RG, and the identification between the extra term in (\ref{twelve}) and the extra NSNS flux, require some identification of $\mu$ with some space-time coordinate. It is in this holographic identification where the gravitational back reaction  enters into the game. We will come back to this issue in section 4. 

The whole discussion until this point can be easily presented for the mirror type B version. In this case we will work on the deformed conifold $X^*$ with $N$ units of RR flux and we will relate $W^{np}(S)$ with the GVW superpotential
\be
W^{np}(S)=\int_{X^*} \Omega \wedge H_{RR}
\ee 
provided we identify $S$ with the complex modulus $t$. The extra term in (\ref{twelve}) can be again reproduced by adding extra NSNS flux, i.e. $\Omega \wedge \left( H_{RR}+\tau_s H_{NSNS}  \right)$, where $\tau_s$ is the type IIB coupling. In section 3 we will work out in detail the GVW superpotential for the deformed conifold paying especial attention to the problem of domain walls.

\subsection{Open String Tadpoles and Warping Factors}

As discussed above, the derivation of the superpotential (\ref{eight}) only uses the non perturbative information of Chern-Simons gauge theory, or its corresponding closed topological string dual on the resolved conifold. In this subsection we discuss the open string interpretation of the second term in (\ref{twelve}), namely the one associated, in the closed string gravity dual, with the extra NSNS flux.

\begin{figure}
\centering
\includegraphics{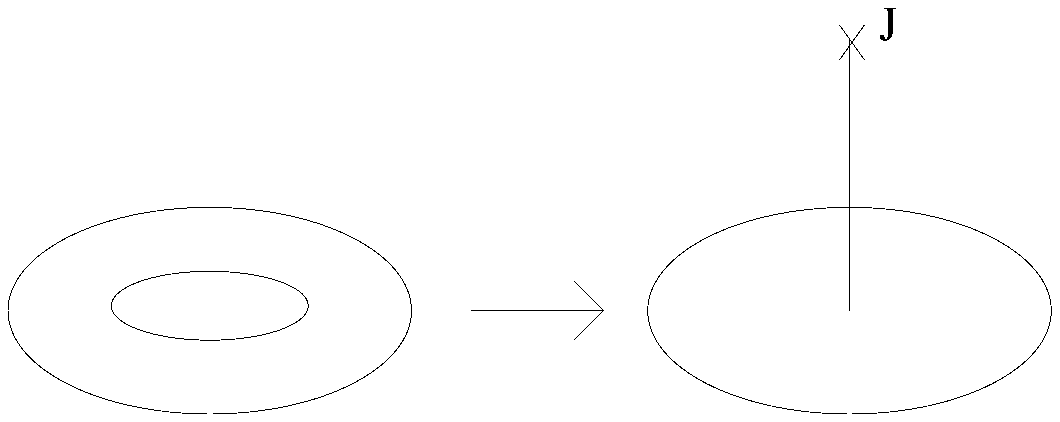}
\begin{center}
\footnotesize{Fig.1}
\end{center}
\end{figure}

Let us come back to consider open strings on $T^*S^3$. A term of type $\tau S$ should formally come (see (\ref{one})) from the amplitude $F^{(s)}_{0,2}$, i.e.  from the annulus. This amplitude can be generically interpreted as a tadpole, i.e. as the one point disk amplitude for a closed string that goes into the vacuum (see fig.1). It is important to stress that this potential tadpole is not part of the perturbative expansion of Chern-Simons \cite{Gopakumar:1998ki}:
\ben\label{free}
F^{(CS)pert}_0(\l )=
\sum_{p=2}^{\infty}\frac{\zeta (2p-2)}{p-1}
\frac{\l^{2p-2}}{2p(2p-1)}
\een
Therefore is something we have not taking into account when we have derived (\ref{eleven}) and its closed string dual.

Before going into more details let us assume that the tadpole exist. The first reaction is of course to try to cancel this tadpole in order to get a consistent string background. A standard way to do that is to use Fischler-Susskind (FS) mechanism \cite{Fischler:ci}, i.e. to modify the background metric. What we are going to suggest next is:
\begin{itemize}
\item[i)] To interpret the extra term $\tau S$ in (\ref{twelve}) as coming from a non vanishing annulus amplitude $F_{0,2}$ tadpole.
\item[ii)] To use FS mechanism as a constructive procedure to derive the gravitational back reaction of NSNS fluxes. More precisely the warping factors. Notice that it is by this procedure how we are going to make contact with an holographic interpretation of the RG.
\end{itemize}

\subsubsection{The Annulus Amplitude}

The annulus amplitude we want to compute is given by \cite{Bershadsky:1993cx}
\be
F_{0,2}=\int_{0}^{\infty}\frac{dL}{L}tr\left( (-1)^F F e^{-LH} \right)
\ee
where the trace is over the states of the open topological string theory 
defined on $T^*S^3$, $L$ is the modulus of the annulus, 
$H$ is the Hamiltonian of the string theory and the insertion of the 
worldsheet fermion number operator $F$ comes from the fact that the theory is twisted. Because 
we are in a topological theory, the massive states correspond to 
pure BRST excitations, so we can ignore them and consider that the trace is 
only over the ground states (g.s.) of the topological string. This leads to
\be
F_{0,2}=\int_{0}^{\infty}\frac{dL}{L}tr_{g.s.} \left( (-1)^F F \right)
\ee
This integral has both ultraviolet and infrared divergences that need a 
regulator. The ultraviolet divergence can be regulated simply by 
introducing a cutoff in the lower limit of integration in the $L$-integral. 
The infrared divergence is regularized by introducing a sort of mass 
gap in the Hamiltonian of the theory, so we get
\be
F_{0,2}=\int^{\infty}_{\left(\mu \a '\right)^{-1}}\frac{dL}{L}NJ e^{-Lm^2 \a '}
\ee
where $NJ=tr_{g.s.} ((-1)^F F)$ is a factor depending on the particular open topological string we are considering. 
Thus, $F_{0,2}$
contribution to the superpotential becomes
\be
\label{tadcon}
2JNS\log \left(\frac{\mu^2}{m^2}\right)
\ee
from which we derive the superpotential
\be
\label{twenty}
W(S)=NS \left[ \log \left( \frac{S}{\m^3}  \right)-1  \right] + 4 NJ \log \left( \frac{\m}{m}  \right)S
\ee
which is just the VY superpotential for $\ene=1$ SYM\footnote{Notice that we are free to identify the energy scale inside the $\log$ in (\ref{eleven}) with the UV regulator of the annulus.}. The tadpole $J$ can be fixed from (\ref{twenty}) by imposing invariance with respect to $\m$. In this way we observe that the {\em stringy} scale $m^{3}$ becomes precisely
the expectation value $<S>$ for the glueball superfield. Recall that the identification of the annulus, on the one hand with the gauge coupling, and on the other hand with the NSNS flux, agrees with the usual relation between $g_{YM}$ and NSNS flux proposed in the context of the gauge/gravity correspondence\footnote{See, for instance, \cite{Imeroni:2003jk} for a nice review}.

\subsubsection{The Fischler-Susskind Mechanism and Holography}

\label{fth}

At this point the reader can wonder on the meaning of the  stringy scale $m$ in standard string theory. Recall that this scale comes from the way we regulate the loop string amplitude $F_{0,2}$ in the IR. As it is well known \cite{Fischler:ci}, loop string tadpoles induce a renormalization of the background metric
\be
\label{renmet}
 g_{\m\n}^0\rightarrow g_{\m\n}^0 + R_{\m\n}^0 \log \left( \frac{\m}{m} \right)  + JN\l_s \log \left( \frac{\m}{m} \right)   g_{\m\n}^0
\ee
where the first term in (\ref{renmet}) comes from the worldsheet sigma model renormalization and $\lambda_s$ is the string coupling constant. The condition of fixing the string beta function
equal to zero leads from (\ref{renmet}) to string backgrounds with a non vanishing  cosmological constant proportional to the tadpole. The scales $\mu$ and $m$ are hidden. On the other hand, here we suggest a different interpretation of the tadpole that consists on treating it in a non critical holographic way \cite{Alvarez:1999cb,Khoury:2000hz}. In this interpretation we associate $\mu$ in (\ref{renmet}) with a holographic coordinate and we use the tadpole to induce a geometrical warping factor.

Let us start with a ten dimensional metric of type
\be
\label{KT}
ds_{10}^2 = h^{-1/2}(\m)dx_{4}^2 + h^{1/2}(\m)ds_{6}^2
\ee
and let us take for the internal six dimensional Calabi-Yau manifold the singular conifold metric \cite{Candelas:js}
\be
ds_{6}^2 =dr^2 + r^2 ds_{T^{1,1}}^2
\ee
Thus
\be
h^{1/2} ds_{6}^2 =h^{1/2}dr^2+h^{1/2}r^2 ds_{T^{1,1}}^2
\ee
Our ansatz is
\be
\label{tadsol}
 h^{1/2}r^2=\a' JN\l_s \log \left( \frac{\m}{m} \right)
\ee
Notice that $r^2h^{1/2}$ defines the geometrical warping of the lagrangian submanifold $S^3$ we are using to define the open topological string amplitudes. Moreover, the ansatz (\ref{tadsol}) defines a relation between the holographic coordinate $\mu$ and $r$.

(\ref{tadsol}) is, in certain sense, not surprising. We can see it by considering the explicit example of the KS solution \cite{Klebanov:2000hb} in the limit in which the deformation of the conifold is small. In that limit the solution is valid far from the apex of the cone, where the gauge group is not $SU(N)$, and can be approximated by the Klebanov-Tseytlin (KT) solution \cite{Klebanov:2000nc}, which has a warping factor
\be
\label{KTwarping}
h^{1/2}_{KT}(U)= \mbox{cte} \frac{\l_sN}{\a' U^2}\left( \log \frac{U}{U_0 e^{-1/4}} \right)^{1/2}
\ee
where $U=\frac{r}{\a'}$ is the holographic coordinate\footnote{Remember that \cite{Klebanov:2000hb} works in the limit $\a' \to 0$, $U$ fixed} and $U_0$ is a regulator related to the deformation. The result (\ref{tadsol}) is exactly the warping factor we could obtain if we were able to extrapolate (\ref{KTwarping}) to the zone $U \simeq U_0$, where the 
gauge group is $SU(N)$,
\be
h^{1/2}_{KT} \sim \frac{\lambda_s N}{\a' U^2}2\log \left( \frac{U}{U_0 e^{-1/4}} \right)
\ee
provided the identifications $\m \sim U$, $m \sim U_0$.

In summary, the moral of this exercise is to show that the gravitational back reaction of NSNS fluxes can be interpreted, once we associate these fluxes with the one loop string tadpoles, as a generalized form of the FS mechanism. In this interpretation the dynamically generated scale of the dual gauge theory is related to the IR cutoff $m$ we have introduced in the regularization of the tadpole.

\section{VY Superpotential and the Deformed Conifold}

In this section we will consider type IIB closed strings on the deformed conifold $X^*$
\be
\sum_{a=1}^{4} z^{2}_{a} =\e^2
\ee
with non vanishing $G_3 =H_{RR}+\t_s H_{NSNS}$ flux. The induced GVW superpotential should be given by
\be\label{superpotential}
W=\int_{X^*}\Omega \wedge (H_{RR}+\t_s H_{NSNS})
\ee

The target of this section is to provide a detailed discussion on the following three issues. First of all the relation between
(\ref{superpotential}) and VY superpotential for $\ene=1$ SYM. Secondly to work out explicitly the different supersymmetric vacua 
and the corresponding domain walls. Finally we will discuss the derivation of beta functions. All of those issues has been 
worked out in references \cite{Gukov:1999ya,Loewy:2001pq,Herzog:2001xk,Cachazo:2001jy,Imeroni:2003cw,Imeroni:2003jk,DiVecchia:2004dg,Merlatti:2004ea}. We will focus our attention on those aspects that we believe require still some further analysis.

\subsection{The Superpotential}

Let us take for the deformed conifold a canonical homology basis of three cycles $( A^0 , A , B^0 , B )$. If we neglect the contribution coming from $A^0$ and $B^0$ we get from (\ref{superpotential})
\be\label{superciclos}
W=\int_A \Omega \int_B G_3 -\int_B \Omega \int_A G_3
\ee
where $\Omega$ is the holomorphic 3-form
\be
\Omega = \frac{1}{4\pi^{2}} \frac{1}{(2\pi \alpha')^{3}} \int_{I=0} \frac{dz_{1}
\wedge dz_{2} \wedge dz_{3}
\wedge dz_{4}}{I}
\ee
with 
\be
I=\sum_{a=1}^{4} z^{2}_{a} -\e^2
\ee
which leads to
\be
\label{omega}
\Omega = - \frac{1}{2\pi i} \frac{1}{(2\pi \alpha')^{3}} \frac{dz_{1} \wedge dz_{2} \wedge dz_{3}}
{2 \sqrt{\epsilon^{2} - z_{1}^{2} - z_{2}^{2} - z_{3}^{2}}}
\ee
In terms of coordinates $(\rho,\psi,\theta_1,\phi_1,\theta_2, \phi_2)$ introduced in appendix B.1, the cycles $A$ and $B$ are
\bea
A: & \rho=\epsilon &   \quad                      \label{Acycle} \\
B: & \psi=0,        & \theta_1 = \theta_2 ,\quad \phi_1=-\phi_2 \label{Bcycle}
\eea

Since the deformed conifold is non compact, the cycle $B$ is non compact. This implies that we need to introduce some cutoff $\rho_c$ 
for the coordinate $\rho^2=\sum_{a=1}^{a=4} z_{a}\bz_a$ in order to compute $W$. In terms of the cutoff $\rho_c$ we define
\be
\phi^{0}= \frac{  \rho_{c}^{2}}{4(2\pi \alpha')^{3} }
\ee
and
\be
\label{phidef}
 \phi = \frac{\epsilon^{2}}{4(2\pi \alpha')^{3}}
\ee
The parameters $\phi$ and $\phi^0$ can be used as projective coordinates on the moduli space of complex structures on $X^*$, thus we can define $t=\frac{\phi}{\phi^0}$ as the complex moduli coordinate.

For the cycles we get (see appendix B.2)
\be
\int_{A} \Omega = 2\pi i \phi
\label{intA}
\ee
and 
\be
\label{intB}
\int_B \Omega = \frac{\partial F}{\partial\phi}=2\phi \sqrt {\left( \frac{\phi^0}{2\phi}  \right)2-\frac{1}{4}} + 2\phi \log \left( \sqrt{\frac{\phi^0}{2\phi}+\frac{1}{2}}+ \sqrt{\frac{\phi^0}{2\phi}-\frac{1}{2}} \right) 
\ee
where the prepotential $F(\phi^0 , \phi )$, given by the genus zero amplitude of the type B topological string theory on $X^*$, is an homogeneous function of degree two which does not depend on the back reaction of the fluxes. Notice that the result (\ref{intB}) is different from the one in \cite{Imeroni:2003jk,DiVecchia:2004dg,Merlatti:2004ea}. The reason is that we have used a different cutoff in the computation of the integral (\ref{intparaB})

After adding $N$ units of RR flux
\be
\label{intAG}
\int_A H_{RR}=N
\ee
we get the superpotential
\be
\label{W}
W=-\frac{\partial F}{\partial \phi}N
\ee
with $\frac{\partial F}{\partial \phi}$ given in (\ref{intB}).

An interesting output of the analysis of section 2 is that, if we want to identify (\ref{W}) with (\ref{eleven}) by using
\be
\label{Vafamirror}
\phi \sim <S>
\ee
which is the mirror of (\ref{Vafaiden}), the cutoff  $\phi^0$ has to play the role of the $\mu$-scale. Of course, in order to make contact with VY superpotential we need to work 
in the region $\phi \ll \phi^0$, in which the deformed conifold is local mirror to the resolved conifold and in which we can neglect the higher powers of $S$ in (\ref{one}). Only in this region we can safely identify $W$,
 as given in (\ref{W}), with the F-term superpotential (\ref{eleven}) of four dimensional $\ene=1$ SYM. In this region, we get from (\ref{intB})
\be
\label{asin}
\frac{\partial F}{\partial\phi}\simeq \phi^0 + \phi \log \frac{2\phi^0}{\phi}
\ee

Let us now add the NSNS flux on cycle $B$:
\be\label{NSflux}
\int_B G_3 = \int_B H_{RR} +\int_B \tau_s H_{NSNS}
\ee
Again, due to non compactness of the $B$ cycle, (\ref{NSflux}) is a function of the cutoff coordinate  $\phi^0$. Let us denote 
(\ref{NSflux}) formally by $\frac{M(\phi^0)}{2 \pi i}$. In these conditions we get 
\be\label{super}
W=-N\frac{\partial F}{\partial\phi}+M(\phi^0)\phi
\ee
The identification $M \sim \tau$ of section 2 is the standard one in the gauge-gravity correspondence\footnote{(\ref{intBG}) can be also obtained by using a probe brane in this background}
\be
\label{intBG}
\int_B G_3 = \int_B H_{RR} + \t_s \int_B H_{NSNS}= \frac{\theta_{YM}}{2 \pi}- \frac{4 \pi i}{g^2_{YM}(\phi^0)}=\frac{M(\phi^0)}{2 \pi i}
\ee
from which we obtain
\be
\label{Wfinal}
W=-\frac{\partial F}{\partial \phi}N+\left( \frac{8 \pi ^2}{g^2_{YM}(\phi^0)}+i \theta_{YM}  \right)\phi
\ee

\subsection{Supersymmetric Vacua and Domain Walls}

Once we have computed the superpotential, we turn our attention to the supersymmetry of this closed string background. Since we have fluxes, in order to preserve $\ene=1$ supersymmetry in four dimensions, the complex structure $t$ must be such that for the corresponding Hodge decomposition $G_{0,3} = G_{1,2} = 0$. These conditions can be easily derived from $W= \partial_t W = 0$. A concrete example is (see appendix \ref{KSsolution}) the Klebanov-Strassler (KS) supersymmetric solution \cite{Klebanov:2000hb}, although we want to remark that this background is not exactly the gravitational dual of pure $\ene=1$ SYM with gauge group $SU(N)$, but the dual of a $SU(N+M)\times SU(M)$ gauge theory which suffers a cascade of Seiberg dualities. At the bottom of the cascade we are left with an $SU(N)$ gauge theory but, in order the cascade steps to be well-separated, one has to impose that $\lambda_s N \ll 1$, which corresponds to a regime at which the supergravity approximation is not valid. The KS solution corresponds to a vacuum of the gauge theory at which the $G_3$-form is of (2,1)-type, and, therefore, $W=0$\footnote{Since $2\pi i\phi\int_B G_3$ tends, in the Klebanov-Strassler solution and for $\frac{\phi^0}{\phi}\gg 1$, to $N \phi \log \frac{2\phi^0}{\phi}$, we observe that the superpotential (\ref{W}) we have obtained is equal to $-N\phi^0$, and then, it does not vanish. We think this happens due to the fact that we neglect the contribution coming from $A^0$ and $B^0$, which has to be proportional to the cutoff $\phi^0$ we have put. In the following, we will neglect the term $\phi^0$ in (\ref{asin})}.

But this is not the whole story. In fact, if we want to preserve supersymmetry, we must impose
\ben
\label{susy}
D_tW = \partial_t W + \partial_t K W = 0
\een
where $K$ is the K\"ahler potential of the moduli space of complex structures on $X^*$
\ben
K =- \log \int_{X^*} \Omega \wedge \bar \Omega
\een 
By solving (\ref{susy}) we get, in the limit of small $t$, the following set of $N$ vacua
\be
\label{solut}
\phi \simeq 2 \phi^0 e^{-\frac{8 \pi^2}{Ng^2_{YM}}-i \frac{\theta_{YM}}{N}+\frac{2 \pi i n}{N} }
\ee
with $n=0,1,...,N-1$. By using (\ref{Vafamirror}) is clear that these $N$ solutions correspond to the $N$ vacua of $\ene=1$ SYM (the $N$ vevs of $S$).

Since in (\ref{solut}) we have obtained 
$N$ different vacua, we expect domain walls with tension given by
\be
\label{tension}
\Delta W = \int_{X^*} \Omega\wedge (G_3^i -G_3^{i+1})
\ee
where $G_3^i$ represent the different three forms at each side of the domain wall. Because $X^*$ is a non compact Calabi- Yau manifold, the cohomology representative of a domain wall, which classify changes of $G_3$ in crossing the domain wall, lives in $H^{3}_{cpt}(\defcon,\entero)$ (cohomology with compact support). The non-vanishing of (\ref{tension}) means that, although $W= \partial_t W = 0$ on one of the vacua, there are other vacua at which $W$ is different from zero. Moreover, since every vacua satisfy the condition (\ref{susy}), a non vanishing value of $W$ is only possible if $\partial_t W$ is also non vanishing. This means that the $G_{3}$ form {\em on different vacua gets non vanishing $(0,3)$ and $(1,2)$ components}. The issue we would like to discuss now is how, by starting with a supersymmetric vacua with $G_3^{0,3}=G_3^{1,2}=0$, we get $(0,3)$ and $(1,2)$ components of $G_3$ when we cross a domain wall. In other words we want to find the relation between 
the different complex structures associated with the $N$ supersymmetric vacua.

To do this, let us take a transformation $\psi \to \psi + \delta\psi$, which is the gravity dual of a $U(1)_R$ transformation in the gauge theory. Clearly, the periods over $A$ are not modified, but the ones over $B$ are. In particular, for $\rho_c \gg \epsilon$, $\int_{B}H_{RR}=\int_{S^2}C_2 \to \int_{S^2}C_2 +\frac{N}{2 \pi}\delta \psi$. Since the geometry far from $\rho=\epsilon$ has the isometry $\psi \to \psi + \delta\psi$, and since $\int_{S^2}C_2$ is a periodic variable with period 1, only the transformations $\delta \psi=-\frac{2\pi n}{N}$, with $n \in \entero$, preserve the background at infinity. This fact is a manifestation of the anomaly in the $U(1)_R$ symmetry. If we take into account the fact that the whole background is invariant under $\psi \to \psi-2\pi$, we conclude that only the subset
\bea
\label{transfo}
\psi \to \psi - \frac{2\pi n}{N} &; & n=0,1,...,N-1
\eea
change the system while preserving the conditions at infinity. These are the transformations that \cite{Loewy:2001pq,Herzog:2001xk} propose to be the ones corresponding to crossing domain walls. This assertion is compatible with our result, because the transformations that lead from one vacua to another one in (\ref{solut})
are equivalent to $z_a \to e^{\frac{-\pi i n}{N}}z_a$, which, at the regime $\frac{\phi^0}{\phi}\gg 1$ we are working, are also equivalent to (\ref{transfo}) (see appendix \ref{deformed}).

As discussed in the appendix \ref{defoKS}, after one of the transformations (\ref{transfo}) is done, and if we introduce the coordinate $\t$ as 
\be
\cosh \t = 1/t
\ee
we get a change in the complex structure
\be
\label{zatransotro}
z_a \to \left(z_a  \right)_{\psi \to \psi + \delta \psi}= z_a\frac{\sinh \left( \t + i\frac{\delta \psi}{2}  \right)}{\sinh \t} - \bz_a \frac{ \sinh\left(i\frac{\delta \psi}{2}\right) }{\sinh \t}
\ee
Then, $G_3$ is no longer a $(2,1)$-form and, therefore, $W\neq 0$ and $\partial_t W \neq 0$. In fact, whereas $\int_A G_3$ does not change under (\ref{transfo}), $\int_BG_3 \to \int_B G_3 + n$. This is equivalent to say that $G_3$ changes in cohomology
\be
\label{coho}
G_3 \to G_3+ n [A]
\ee
(notice that $[A]$ is the only generator of $H^3_{cpt}(\defcon,\entero)$) which implies that the superpotential changes into
\be
W  \to W +n\int_A \Omega
\ee
The tension of the domain wall is therefore proportional to $\phi$, as expected. 

\subsection{The Renormalization Group}

We are now located at a point suitable on order to clarify an issue concerning the papers \cite{Imeroni:2002me,Imeroni:2003jk,DiVecchia:2004dg}, where a computation of the UV running of $g_{YM}^2$ for {\em pure} $\ene=1$ Super Yang-Mills is carried out from the KS solution. The question is why they obtain the correct running by working in a regime $(\lambda_s N \ll 1)$ at which the warping factor makes the solution to be not valid.

Let us first summarize the approach we have follow in order to get information about beta functions of $\ene=1$ SYM from 
the closed string gravity dual. We first use the deformed conifold with $N$ units of RR flux. We introduce a cutoff $\phi^0$ 
in the computation of the superpotential and we work in the regime $\phi \ll \phi^0$ where the deformed conifold is local mirror 
of the resolved conifold. In this regime powers of $S$ in the superpotential  (\ref{one}) are negligible and we expect to have $\ene=1$ SYM with $N$ 
the rank of the gauge group. Next we add NSNS fluxes that we interpret as 
\be\label{coupling}
\frac{8\pi^2}{g^2_{YM}}=\frac{2\pi}{\lambda_s}\int H_{NSNS}
\ee
The running for $g^2_{YM}$ is obtained by impossing to the superpotential invariance with respect to the cutoff  $\phi^0$. Notice that 
in this approach we do not need any information on the gravitational backreaction of NSNS fluxes or equivalently on the geometrical 
warping factors. By this procedure we get 
\be
\label{gyflujo}
\frac{8\pi^2}{g^2_{YM}}=N\log \frac{2\phi^0}{\phi}
\ee
which is the one loop beta function once we use the identification
\be
\frac{\phi^0}{\phi}= \frac{\mu^3}{\Lambda^3}
\ee
which appear naturally in the above discussion. By this procedure we get from the superpotential the expectation value 
\be\label{vevS}
|<S>|=\mu^3 \exp -\frac{8\pi^2}{g^2_{YM}N}
\ee
which is the correct value if we think of $g^2_{YM}$ in (\ref{vevS}) as the Wilsonian coupling constant $g_W^2 (\mu)$ of (\ref{exactbeta})\footnote{Taking into account more powers of $t$ only gives contributions to the Wilsonian beta function (\ref{one}) that goes like powers of $e^{-\frac{8 \pi^2}{Ng^2_{YM}}}$}. Thus we must 
interpret (\ref{coupling}) as 
\be
\frac{8\pi^2}{g^2_{W}}=\frac{2\pi}{\lambda_s}\int H_{NSNS}
\ee
For the effective coupling constant the right value for $<S>$ is 
\be
|<S>|=\mu^3 \frac{1}{Ng_{eff}^2}\exp -\frac{8\pi^2}{g^2_{eff}N}
\ee
which leads to the NVSZ exact beta function \cite{Novikov:uc}. Quite surprisingly, in \cite{Imeroni:2002me,Marotta:2002ns,Imeroni:2003jk,DiVecchia:2004dg} the exact NVSZ beta function for pure $\ene=1$ SYM is also derived from the KS gravitational background. 

Obiously we can always look for a change of variables $\mu\to \phi^0 (\mu)$ such that
\be\label{rela}
g_{eff}^2 (\mu) = g_W^2 (\phi^0 (\mu))
\ee
Defining $\frac{\phi^0}{\phi}=\cosh \t$ the map $\phi^0 (\mu)$ solving (\ref{rela}) is given by\footnote{This relation was first proposed in \cite{Imeroni:2002me} inspired by \cite{DiVecchia:2002ks}}
\be
\frac{\t}{2\sinh \t}=\frac{\Lambda^3}{\mu^3}
\ee
The change of variables  $\mu\to \phi^0 (\mu)$ must be the holographic analog of the change from holomorphic to canonical variables 
in the gauge dual theory.

\section*{Acknowledgements}

This work was partially supported by  Plan Nacional de Altas 
Energ\'\i as, Grant FBA2003-02-877. 
The work of S.M. is supported by the Ministerio de Educaci\'on y 
Ciencia through FPU Grant 
AP2002-1386. The work of P.R. is supported by a FPU-UAM Grant.

\newpage

\appendix{\Large {\Large {\bf{Appendix}}}}

\begin{appendix}

\section{Topological Free Energy in presence of RR Flux}


\label{teorema}
\renewcommand{\theequation}{A.\arabic{equation}}
\setcounter{equation}{0}

In this appendix we will sketch the proof of the independence of the topological free energy when we turn on RR fluxes. We will do that in a $\hat{c}=5$ model that, for the set of amplitudes that 
we are considering, is equivalent to the critical RNS superstring
\cite{Berkovits:1994wr,Berkovits:1994vy,Berkovits:2003pq}.

The action for the topological model is 
\be
\int d^2z(p_\a\bar{\pd}\t^\a+\bar{p}_\a\pd\bar{\t}^\a+
\frac{1}{2}\pd x^m \bar{\pd}x_m)+S_c
\ee
where $x^m$ are bosons and $p_\a , \t^\a $ are space-time fermions 
(in four dimensions). $S_c$ is the action for a superconformal 
field theory associated with a Calabi-Yau six dimensional manifold.

This formulation presents a $N=2$ (twisted) superconformal symmetry.
The generators can be written 
\bea
T &=& p_\a\pd\t^\a+ \frac{1}{2}\pd x^m \pd x_m + T_c\\
G^+ &=& \t_\a\pd x^{\a \dot{+}}+G^+_c\\
G^- &=& p_\a\pd x^{\a \dot{-}}+G^-_c\\
J &=& \t^\a p_\a + J_c
\eea
for the left-moving side and the same but using barred variables 
for the right-moving side.

For our purposes we need two kinds of vertex operators
\begin{enumerate}
\item Graviphoton vertex operators
\be 
R(x,\t ,\bar{\t})
\ee
where
\be
F_{\a\b}=\pd_{\a \dot{+}}\pd_{\b \dot{+}}R|_{\t =\bar{\t}=0}
\ee
is the self-dual graviphoton field strength and
\be
R_{\a\b\g\d}=\pd_{\a \dot{+}}\pd_{\b \dot{+}}D_{\g}\bar{D}_{\d}R|_{\t =
\bar{\t}=0}
\ee
is the self-dual Riemann tensor
\footnote{$\pd_{\a\dot{\b}}= \s^{m}_{\a\dot{\b}}\pd_m$ as usual}. Note that this vertex have no Calabi-Yau dependence and that the four 
dimensional part of the model is simply the $\hat{c}=2$ 
self-dual string.
\item Chiral vertex operators
\be
\Phi(x,\t ,\bar{\t})\s
\ee
$\s$ (in the IIA case)  is an element of $G^{+}_c$ cohomology with $+1$ $U(1)_c$ charge. 
$\s$ can be associated with a cycle in $H_{(1,1)}(X)$ where $X$ is the 
Calabi-Yau. $\Phi$ live in the $G^{+}_{4d}$ cohomology. $\Phi|_{\t =\bar{\t}=0}=t$ is a chiral modulus of the Calabi-Yau, and 
$D_\a \bar{D}_\b \Phi |_{\t =\bar{\t}=0}$ is the self-dual RR flux 
associated with the modulus. Note that because we are considering chiral vertex operators associated with elements of 
$H_{(1,1)}(X)$, this correspond to turn on RR flux for the two-form of IIA. This last identification is quite 
technical and comes from the precise relation of the formalism that we 
are considering and the standard RNS string (see \cite{Berkovits:1995cb} for precise 
details).
\end{enumerate}

On this model we have that genus $g$ scattering amplitudes take the form
\be
A_{g,M}= \langle|\int m_1 \int \m_1G^- \dots 
\int m_{3g-3+M} \int\m_{3g-3+M}G^-|^2 V_1\dots V_M \rangle
\ee
where $\m$ are the Beltrami differentials and $m$ the moduli parameters. 
Looking at the scattering amplitudes you can see that for $U(1)$ 
charge conservation we have the selection rule
\be
5(1-g)=J_1 +\dots +J_M -(3g-3+M)
\ee
in order to have a scattering amplitude different from zero. $J_i$ 
are the different $U(1)$ charges of the $M$ vertex operators inserted, so
\be
2g-2+M=J_1+\dots + J_M 
\ee
is valid for left and right $U(1)$ charges.

Let us consider amplitudes with $M-2g$ chiral vertex operators inserted and $2g$ 
of the graviphoton type. This implies that 2 units of $U(1)$ charge 
must come from the 4d sector. The form of the 4d part of the $U(1)$ current  implies that the charge 
comes from $\t\,\bar{\t}$ zero modes. Let us choose $2g$ of the Beltrami differentials and moduli as the locations 
of the graviphoton vertex operators. Then amplitudes take the form
\bea
A_{g,M}= \langle|\int m_1 \int\m_1G^-_c \dots 
\int m_{g-3+M} \int\m_{g-3+M}G_c^-|^2\\\nonumber 
\Phi_1\s_1\dots\Phi_{M-2g}\s_{M-2g}
\int d^2z_1W_1\dots\int d^2z_{2g}W_{2g}  \rangle
\eea
where 
\be
W=\int G^-\int \bar{G^-} R(x,\t,\bar{\t})
\ee
Note that $(3-3g)$ $U(1)$ charges are needed for the internal sector so only 
$G^-_c$ contributes in $\int \m G^-$. Note also that the vertex 
operators we are considering here are chiral primaries of the $\hat{c}=5$ 
theory (living in $G^+$ cohomology and having only single poles with $G^-$).

The condition of chiral primarity for a general vertex operator is that 
$\s$ have to be a primary of the 6d theory and
\bea
\frac{\pd}{\pd\t_\a}\pd_{\a\dot{+}}\Phi = 0\\
\frac{\pd}{\pd\t_\a}\pd_{\a\dot{+}}R = 0
\eea
But this condition is not needed in the amplitudes considered above 
because only the 6d part of $G^-$ contribute due to the selection rule.
The condition of ``decoupling'' of the 4d part of $G^-$ is also necessary 
to prove the space-time supersymmetry of the amplitude.

We want now to extract the corresponding 4d F-terms associated with 
this amplitude. In order to 
do this we have to integrate over the zero modes of the $x,\t,\bar{\t}$ 
fields and use the graviphoton vertex to absorb the $p$ zero modes.  We need to remember that the $\t\bar{\t}$ component of $R$ is 
the self dual graviphoton field strength $F_{\a\b}$. By using this, the 
amplitudes take the form
\bea
A_{g,M}= \int d^4 xd^2\t d^2\bar{\t}\Phi_1\dots\Phi_{M-2g}
F_1\dots F_{2g}\\\nonumber
\langle|\int m_1 \int\m_1G^-_c \dots 
\int m_{g-3+M} \int\m_{g-3+M}G_c^-|^2 
\s_1\dots\s_{M-2g}  \rangle_c
\eea
so the associated F-terms look
\be
f_{i_1\dots i_{M-2g}}\int d^4 xd^2\t d^2\bar{\t}\Phi_1\dots\Phi_{M-2g}
F_1\dots F_{2g}
\ee
where $f$ depends only of the $N=2$ , $\hat{c}=2$ theory. 
This is an example of factorization of the amplitudes into 4d and 
6d parts. 
If the 6d amplitudes can be defined in terms of a prepotential that 
depends of the K\"ahler moduli (in the IIA case) or the complex structure 
moduli (in the IIB case) $F_g(t_i)$ then
\be
f_{i_1,\dots i_{M-2g}}=\pd_1\dots\pd_{M-2g}F_g
\ee

In conclusion we have obtained that by turning on the RR flux on 2-cycles of the Calabi-Yau (so turning on the RR 2-form of the IIA superstring) we 
generate F-terms in the four dimensional effective action that can be computed in terms of derivatives of the topological string  
amplitudes on the Calabi-Yau. This proves the independence of the topological amplitudes 
when we turn on 2-form fluxes. The same thing works for NSNS fluxes associated with the two form (that appears as imaginary parts of the 
IIA superstring 2-form). In the case of IIA four forms and six forms, the F-term cannot be written in terms of derivatives of the topological amplitude (see in \cite{Vafa:2000wi} the concrete structure of the F-terms associated to these kind of fluxes). Therefore, turning on this 4-form and 6-form fluxes does not modify the F-term generated by the 2-form.

\section{ Deformed Conifold Geometry and Klebanov-Strassler Solution}


\label{defoKS}
\renewcommand{\theequation}{B.\arabic{equation}}
\setcounter{equation}{0}


\subsection{U(1) Transformations on the Deformed Conifold}

\label{deformed}

The deformed conifold $X^*$ is a non-compact Calabi-Yau threefold defined by \cite{Candelas:js}
\be
\sum_{a=1}^{a=4} z_{a}^{2} = \epsilon^{2}
\ee
which is smooth provided that $\epsilon \neq 0 $\footnote{We take $\epsilon$ to be real}. Differently from its singular limit $X^{||}$, given by $\epsilon =0$, $\defcon$ does not have the continuous symmetry $U(1):z_a \to e^{i\frac{\Delta \psi}{2}}z_a$, with $\frac{\Delta \psi}{2} \in [0,2\pi)$. The only subgroup that survives after setting $\epsilon \neq 0$ is ${\entero}_2$, generated by $\frac{\Delta \psi}{2}=\pi$.

By splitting $z_a$ into its real and imaginary parts $z_a=x_a+iy_a$ one can see that $X^* \approx T^{\ast} S^3$, where the 3-cycle $S^3$, which we call $A$, is given by $y_a=0$. We take as the non-compact coordinate
\be
\label{rho}
\rho^2=\sum_{a=1}^{a=4} z_{a}\bz_a
\ee
Then $\rho \geq \epsilon$. The sections $\rho = cte \neq \epsilon$ have the topology $S^3 \times S^2$. Another useful coordinate is $\t$, given by
\be
\cosh \t = \frac{\rho^2}{\epsilon^2}
\ee
For the whole Calabi-Yau we use the parametrization
\be
\left[
\ba{cc}
z_3+iz_4 & z_1-iz_2\\
z_1+iz_2 & -z_3+iz_4
\ea
\right]=\left[
\ba{cc}
a & -\bar{b}\\
b & \bar{a}
\ea
\right]\left[
\ba{cc}
0& \epsilon \cdot e^{\frac{\t}{2}}\\
\epsilon \cdot e^{-\frac{\t}{2}} & 0
\ea
\right]\left[
\ba{cc}
k & -\bar{l}\\
l & \bar{k}
\ea
\right]
\ee
where
\be
\left\{
\ba{c}
a=\cos \frac{\theta_1}{2} e ^{\frac{i}{2}(\psi_1+\phi_1)}\\
b=\sin \frac{\theta_1}{2} e ^{\frac{i}{2}(\psi_1-\phi_1)}\\
k=\cos \frac{\theta_2}{2} e ^{\frac{i}{2}(\psi_2+\phi_2)}\\
h=\sin \frac{\theta_2}{2} e ^{\frac{i}{2}(\psi_2-\phi_2)}
\ea
\right.
\ee
having, in addition to (\ref{rho}), the coordinates
\bea
\theta_i \in [0,\pi),\qquad i=1,2\\
\phi_i \in [0,2\pi),\qquad i=1,2\\
\psi=\psi_1+\psi_2 \in [0,4\pi)
\eea
since $\psi_i \in [0,2\pi)$.
It is easy to see that, whereas for $X^{||}$ the transformation $z_a \to e^{i\frac{\Delta \psi}{2}}$ is the same as the transformation $\psi \to \psi + \delta \psi$, provided that $\Delta \psi = \delta \psi$, for $X^*$ the relation between both transformations is no longer straightforward. This implies in particular that, under
\be
\label{psitrans}
\psi \to \psi + \delta \psi
\ee
all the $z_a$ changes in the way
\be
\label{zatrans}
z_a \to \left(z_a  \right)_{\psi \to \psi + \delta \psi}= z_a\frac{\sinh \left( \t + i\frac{\delta \psi}{2}  \right)}{\sinh \t} - \bz_a \frac{ \sinh\left(i\frac{\delta \psi}{2}\right) }{\sinh \t}
\ee
One can see that after this transformation is done, the $A$-cycle, which is given by (\ref{Acycle}), is the same as before, whereas the $B$-cycle (\ref{Bcycle}) changes to
\be
B: \qquad \psi=-\delta \psi , \qquad \theta_1=\theta_2 , \qquad \phi_1=-\phi_2
\ee


\subsection{Computation of the Periods}

\label{compu}

For the $A$-cycle we have
\bea
\int_A \Omega &=& -\frac{1}{4 \pi i}\frac{1}{(2 \pi \alpha')^3}\int_A \frac{dz_1 \wedge dz_2 \wedge dz_3}{\sqrt{\epsilon^2-z_1^2-z_2^2-z_3^2}}=\nonumber \\
&=&-\frac{1}{4 \pi i}\frac{1}{(2 \pi \alpha')^3} \left[  \int_{R \in [-\epsilon,0]} \frac{dx_1 \wedge dx_2 \wedge dx_3}{-\sqrt{\epsilon^2-R^2}} +\int_{R \in [0,\epsilon]} \frac{dx_1 \wedge dx_2 \wedge dx_3}{\sqrt{\epsilon^2-R^2}}\right]= \nonumber \\
&=& 2\pi i \phi
\eea
where $R=\sqrt{x_1^2+x_2^2+x_3^2}$.

On the other hand, since (\ref{Bcycle}) implies that
\be
B: \qquad y_1=0,\qquad x_2=0,\qquad y_3=0, \qquad y_4=0
\ee
we have
\bea
\int_B \Omega &=& \frac{1}{4 \pi i}\frac{1}{(2 \pi \alpha')^3}\int_{R^\prime \in \left[0,\sqrt{\frac{\rho_c^2-\epsilon^2}{2}}\right]} \frac{dx_1 \wedge dx_3 \wedge dx_4}{\sqrt{\epsilon^2-x_1^2-x_3^2-x_4^2}}=\nonumber \\
&=&2\phi \sqrt {\left( \frac{\phi^0}{2\phi}  \right)^2-\frac{1}{4}} + 2\phi \log \left( \sqrt{\frac{\phi^0}{2\phi}+\frac{1}{2}}+ \sqrt{\frac{\phi^0}{2\phi}-\frac{1}{2}} \right) \label{intparaB}
\eea
where $R^\prime=\sqrt{x_1^2+x_3^2+x_4^2}$.


\subsection{$G_3$ Flux in Klebanov-Strassler Solution}

\label{KSsolution}

The Klebanov-Strassler solution \cite{Klebanov:2000hb} is the closed string background ${\real}^{1,3} \times X^*$ warped due to fluxes coming from $N$ D5-branes, which we had on the open string side of the duality, wrapped on the collapsing $S^2$ of the conifold.
\bea
ds^2& =&  h^{-1/2} ( \tau ) \eta_{\alpha \beta}dx^{\alpha} d x^{\beta} +
h^{1/2} (\tau ) ds_{X^*}^{2} \label{dsKS}\\
H_{NSNS}& =& \frac{\l_s N }{8\pi^2} d \left[\left( f (\tau) ( g^1 \wedge g^2 )
+ k (\tau) (  g^3 \wedge g^4 )\right) \right] \\
H_{RR}& =& \frac{N}{8 \pi^2} \left[ \frac{1}{2}g^5 \wedge \left( g^1 \wedge g^2 + g^3 \wedge g^4\right)+
 d \left( \left(F (\tau)-\frac{1}{2}\right) ( g^1 \wedge g^3 +  g^2 \wedge g^4 ) \right) \right]
 \label{HRRKS}\\
{\tilde{F}}_5 &=& {\cal{F}}_5 + {}^* {\cal{F}}_5 ~~~,~~~{\cal{F}}_5 =
\frac{\l_s N^2}{(8 \pi^2)^2} {\ell} (\tau ) g^1 \wedge g^2 \wedge 
g^3 \wedge  g^4 \wedge  g^5 ~ \label{f5KS}\\
e^\Phi &=&\l_s
\eea
$ds_{X^*}^{2}$ is the deformed conifold Ricci-flat K\"ahler metric \cite{Candelas:js,Klebanov:2000hb} and
\bea
F (\tau) = \frac{\sinh \tau - \tau}{2 \sinh \tau}~,~
f (\tau) =\frac{\tau \coth \tau -1 }{2 \sinh \tau} ( \cosh \tau -1)
\label{FtauKS}
\\
k(\tau ) = \frac{\tau \coth \tau -1 }{2 \sinh \tau} ( \cosh \tau +1)~,~
\ell (\tau ) = \frac{\tau \coth \tau -1 }{4 \sinh^2 \tau}
(\sinh 2 \tau -2 \tau )
\label{ktauKS}
\\
h (\tau ) = (\l_s N \alpha' )^2 2^{2/3}
\epsilon^{-8/3} \int_{\tau}^{\infty} dx \frac{ x \coth x -1 }{ \sinh^2
  x} (\sinh 2 x -2 x )^{1/3}
\label{htauKS}
\eea
De 1-forms $g^i$ are defined by
\be
g^1 = \frac{e^1-e^3}{\sqrt{2}}~,~g^2 = \frac{e^2-e^4}{\sqrt{2}}~,~
g^3 = \frac{e^1+e^3}{\sqrt{2}}~,~g^4 = \frac{e^2+e^4}{\sqrt{2}}~,~
g^5 = e^5
\label{gforms}
\ee
where
\bea
e^1 = -\sin\theta_1 d\phi_1~,~e^2 = d\theta_1~,~
e^3 = \cos\psi \sin\theta_2 d\phi_2 - \sin\psi d\theta_2 \nonumber\\
e^4 = \sin\psi \sin\theta_2 d\phi_2 + \cos\psi d\theta_2~,~
e^5 = d\psi + \cos\theta_1 d\phi_1 + \cos\theta_2 d\phi_2
\label{eforms}
\eea
For this solution $G_3=H_{RR}-\frac{i}{\l_s}H_{NSNS}$ has periods
\bea
\label{intAKS}
\int_A G_3 & =  \int_A H_{RR} & =  N \\
\label{intBKS}
\int_B G_3 & =  \int_B H_{RR}& -\frac{i}{\l_s} \int_B H_{NSNS}  =  \frac{N}{2\pi}c+\frac{N}{2 \pi i}\left[ f(\t_c)+g(\t_c)  \right]
\eea
being $B$ the compactified $B$-period of (\ref{Bcycle}) and $c$  an arbitrary constant. As it is proved in \cite{Cvetic:2000mh}, $G_3$ is a $(2,1)$-form. Expressed in terms of the obvious 1-forms on the deformed conifold: $dz_a$, $d\bz_a$, $a=1,2,3,4$ \cite{Herzog:2001xk}:
\bea
\label{GKS}
G_3=\frac{N}{8\pi^2 \sinh ^4 \t} \left[  \frac{\sinh (2\t)-2\t}{\sinh \t}\left(  \epsilon_{abcd}z_a\bz_bdz_c\wedge d \bz_d \right)\wedge \left(\bz_edz_e   \right) +\right. \nonumber\\
\left. +2 \left( 1-\t \coth\t   \right)\left( \epsilon_{abcd}z_a\bz_bdz_c\wedge dz_d   \right)\left( z_ed\bz_e  \right) \right]
\eea

Let us analyze the consequence of changing the $2$-cycle where the initial $D5$-branes are wrapped, in such a way that for that cycle $\psi=\psi_0$ instead of being zero. The Klebanov-Strassler solution would be modified, in the sense that all the forms $g^i$ that appear in the fluxes change
\be
g^i \to \left(g^i\right)_{\psi \to \psi - \psi_0}
\ee
One can see that, after the transformation is done, (\ref{intAKS}) does not change, but (\ref{intBKS}) does
\be
\int_B H_{RR}=\frac{N}{2\pi}\left[ c- \psi_0 - \left( 2F(\t_c)-1  \right)\sin \psi_0   \right]
\ee
where we have taken for $B$ the cycle (\ref{Bcycle}). By applying (\ref{zatrans}) with $\delta \psi=-\psi_0$ to (\ref{GKS}), it is easy to see that the transformed solution has a  $G_3$ which picks up $(0,3)$-form, $(3,0)$-form and $(1,2)$-form components that, although are negligible at large $\t$, are different from zero.

\end{appendix}

\end{document}